\documentclass[12pt]{article}
\usepackage[T2A]{fontenc}
\usepackage[cp866]{inputenc}
\usepackage[english]{babel}
\usepackage{amsthm,amssymb,amsmath}

\numberwithin{equation}{section}
\def\beq{\begin{equation}}
\def\eeq{\end{equation}}

\begin{document}

\title{\bf First Passage Time Distribution and Number of Returns
for Ultrametric Random Walk}
\author{
V.A. Avetisov, \thanks{Institute of Chemical Physics, RAS, Moscow,
e-mail: avetisov@chph.ras.ru} \\
A. Kh. Bikulov,\thanks{Institute of Chemical Physics, RAS, Moscow,
e-mail: bikulov1903@rambler.ru}\\ and\\ A. P. Zubarev \thanks
{Samara State University of Architecture and Civil Engineering, Samara, e-mail: apzubarev@mail.ru}}
\date {$~$}
\maketitle

\begin {abstract}
In this paper, we consider a homogeneous Markov process $\xi
(t;\omega )$ on an ultrametric space  $Q_{p} $ , with
 distribution density $f(x,t),\; x\in Q_{p}$, \mbox{$t\in R_{+} $},
satisfying the ultrametric diffusion equation $ \frac{\partial
}{\partial t} f(x,t)=-D_{x}^{\alpha } f\left(x,t\right). $
  We construct and examine a random variable $\tau
_{Z_{p} } (\omega )$ that has the meaning the first passage times.
Also, we obtain a formula for the mean number of  returns on the
interval $ (0,t]$ and give its asymptotic estimates for large $t$.
\end {abstract}


\section*{Introduction}

Ultrametric random processes and their physical and biological  applications have
recently been attracting much attention, especially in connection with
modelling the dynamics and evolution of complex  systems
  characterized by multidimensional rugged energy  landscapes (fitness landscapes)
with a huge number of local minima (see, for instance,
\cite{OS}--\cite{AZ}). It is clear that a description of the
dynamics on such landscapes requires adequate approximations. As
shown recently, a reasonable approximation  for the dynamics of some
biological systems (in particular, proteins) can be chosen in the
form of random ``jumps'' between local minima of a landscape, under
the assumption that  the only key factor is the maximal activation
barrier on the landscape that separates these local minima
\cite{BK}. In this case, the local minima are clustered in
``basins'' of minima hierarchically embedded in one another.
Accordingly,  the dynamics of such a system is described in terms of
random ``jumps'' between the basins. As shown in recent publications
\cite{ABK}--\cite{ABO} such approximations can be naturally
described in terms of ultrametric random processes, and it turns out
that the $ p$-adic pseudodifferential equation of ultrametric
diffusion (introduced in \cite{VVZ} and called there \textit{the
equation of Brownian motion on the $ p$-adic line}) gives an
adequate
 description of protein dynamics \cite{ABKO},\cite{AB}.

Clearly, a  physically meaningful application of the
ultrametric diffusion equation requires an answer to  some questions
pertaining to  the description of experimentally  observable quantities.
 In this connection, it should be
mentioned that the observable quantity in a real experiment
corresponds to some specific  degrees  of freedom (determined by the
adopted method of measurement) of a complex system with many degrees
of freedom. In some cases, the observable quantity corresponds to a
characteristic of the system averaged over all its states
 \cite{ABKO}, in other situations, it is determined  by some
specific states \cite{AB}. In a situation of this kind, there arise
classical probability problems for  ultrametric diffusion such as
the problem  of the first passage time distribution and the problem
of the number of returns.

In this paper, we consider a homogeneous Markov process $\xi
(t;\omega )$ on an ultrametric space  $Q_{p} $ (ultrametric random
walk), with
 distribution density $f(x,t),\; x\in Q_{p}$, \mbox{$t\in R_{+} $},
satisfying the equation
$$
\frac{\partial }{\partial t} f(x,t)=-D_{x}^{\alpha } f\left(x,t\right),
$$
 usually called \textit{the ultrametric
diffusion equation} (for definitions and notation see below). We
consider a specific random process  $\xi (t;\omega )$, namely, that
whose  distribution density satisfies the Cauchy problem for the
ultrametric diffusion equation  with the initial density in a
 domain   $Z_{p} \subset Q_{p} $.

Our aim is to construct and examine a random variable   $\tau _{Z_{p} } (\omega )$
that has the meaning the first time instant when
the trajectories of the random process
  $\xi (t;\omega )$ return to the domain  $Z_{p} $.
To study this problem, we first prove that
the distribution density of   $\tau _{Z_{p} } (\omega )$, denoted by $ f(t)$,
 satisfies a nonhomogeneous Volterra
equation, then we construct a solution of that equation and examine its properties.
On the other hand, we show that the first passage time distribution density
can be represented as a functional of a density function which is a solution of the
ultrametric diffusion equation with the absorbing region
$Z_{p} $.  It is shown that these two approaches are equivalent.
In the last part of the paper, we consider the problem of the number of returns
to the domain $ Z_p$
on the time interval   $(0,t]$. We obtain a recurrent equation for
 the probability   $q^{(m)} (t)$ of the   $m$-th return  on the time interval $ (0,t]$,
 as well as a recurrent equation for the probability   $h^{(m)} (t)$
 of precisely  $m$ returns on the interval  $(0,t]$.
 We study the properties of the functions
$q^{(m)} (t)$,  $h^{(m)} (t)$ and obtain a formula for
the mean
number of  returns  on the interval $ (0,t]$ and give its asymptotic estimates
for large   $t$.

 Section 1 contains some basic  facts from $ p$-adic analysis and the theory of
 random processes. These facts  are  used for  the introduction of the necessary
 notation and definitions. In Section~2, we consider the
 first passage problem for  ultrametric random  walk.
In Section~3, we introduce and examine a $ p$-adic analogue of the
diffusion  equation with an absorbing region  for the first passage problem.
 Section~4 is dedicated to the problem of the number of returns for
 ultrametric diffusion.

\section{Elements of $p$-Adic Analysis and the Theory of \\ Random Processes
}

Let   $Q$ be the field of rational numbers and   $p\in Q$ a fixed prime.
Any rational number  $x\ne 0$  can be uniquely represented in the form

$$
 x=p^{\gamma } \frac{a}{b},
$$
where  $a,{\rm \; }b,{\rm \; }\gamma \in Z$ are integers;  $a$  and $ b$ are coprime
positive integers indivisible by   $p$. The \textit{$ p$-adic norm\/}
$\left|x\right|_{p} $ of   $x\in Q  $ is defined by the relations
 $\left|x\right|_{p} =p^{-\gamma } $,  $\left|0\right|_{p} =0$.
 The completion of the field of rational numbers   $Q$
 with respect to the   $p$-adic norm is denoted by $ Q_p$ and is called
 \textit{the field of $ p$-adic numbers}. The set $ Q_p$
 endowed with the  metric  $\rho
(x,y)=\left|x-y\right|_{p} $ is an ultrametric space which is
 complete, separable, totally disconnected, and
locally compact.  There is a unique (to within a coefficient) Haar measure
$d_{p} x$ on  $Q_{p} $ which is translation-invariant: $d_{p}
\left(x+a\right)=d_{p} x$. We normalize this measure by the condition
$$
\displaystyle \int _{Z_{p} }d_{p} x =1 ,
$$
where  $Z_{p} =\left\{x\in Q_{p} :{\rm \; \; }\left|x\right|_{p}
\le 1\right\}$ is the ring    $p$-adic integers.  There is only one
 measure  $d_{p} x$ satisfying the above condition.

We introduce a class   $W^{\alpha } $   $(\alpha \ge 0)$
of complex-valued functions  $\varphi (x)$ on  $Q_{p} $ satisfying the following conditions:
\begin{itemize}
\item[(i)]
  $\left|\varphi (x)\right|\le
C\left(1+\left|x\right|_{p}^{\alpha } \right)$, where   $C$ is a   constant;

\item[(ii)] there is an integer  $N=N(\varphi )>0$ such that
$\varphi \left(x+x'\right)=\varphi \left(x\right)$ for any  $x\in Q_{p} $
and any  $x'\in Q_{p} $ such that
$\left|x'\right|_{p} \le p^{-N} $.
\end{itemize}

A function  $\varphi (x)$ satisfying these two conditions
is called \textit{locally constant}, and $N(\varphi )$
is called its \textit{index of  locally constancy}.
For a function $ \varphi$ that additionally depends on a real  parameter  $ t$,
we say that  $\varphi \in W^{\alpha } $ \textit{uniformly in\/} $t$, if the constant  $C$
and the index  $N$ do not depend on  $t$ .

Functions in  $W^{0} $  with compact support are called \textit{test functions}
(or \textit{Bruhat--Schwartz functions\/}). The set of all test functions
is denoted by    $D$, and the set of distributions on $ D$  is denoted by $ D'$.

Let  $\chi $ be a normalized additive character of the field
$Q_{p} $. Then  $\chi \in W^{0} $. The \textit{Fourier transform\/} of a function
 $\varphi \left(x\right)\in L^{1} \left(Q_{p} ,d_{p}
x\right)$ is defined by
\beq
\tilde{\varphi }\left(k\right)=\int _{Q_{p} }
\chi \left(kx\right)\varphi \left(x\right)d_{p} x  ,\quad
k\in
Q_{p}  . \label{(1.1)}
\eeq
For  $\tilde{\varphi }\left(k\right)\in L^{1} \left(Q_{p}
,d_{p} k\right)$, the \textit{inverse Fourier transform\/}  is defined by
\beq
\varphi \left(x\right)=\int _{Q_{p} }\chi \left(-kx\right)\tilde{\varphi }
\left(k\right)d_{p} k  ,\quad         x\in Q_{p} . \label{(1.2)}
\eeq

The operator  $D_{x}^{\alpha } $  (\textit{the Vladimirov
pseudodifferential operator}  \cite{VVZ}),  $\alpha >0$,  is defined
on functions
 $\varphi \in W^{\beta }$, $ 0\le \beta <\alpha $,
by the formula
\beq
 D_{x}^{\alpha } \varphi \left(x\right)=-\frac{1}
{\Gamma \left(-\alpha \right)} \int _{Q_{p} }d_{p}
y\frac{\varphi \left(y\right)- \varphi
\left(x\right)}{\left|x-y\right|_{p}^{\alpha +1} } \,  , \label{(1.3)}
\eeq
where   $\Gamma
_{p} (-\alpha )=\frac{1-p^{-\alpha -1} }{1-p^{\alpha } } $  is the $ p$-adic
analogue of the gamma-function.

Below, we consider  random processes over the field
  $Q_{p}$. According to the Kolmogorov axioms , a \textit{measurable space\/}
  is a pair   $\left\{\Omega ,\Sigma \right\}$, where  $\Omega $
  is a set and   $\Sigma $ is a   $\sigma $-algebra of subsets of  $\Omega $.
A \textit{probability space\/} is a triplet
  $\left\{\Omega
,\Sigma ,{\rm P} \right\}$, where   $\left\{\Omega ,\Sigma
\right\}$ is a measurable space and ${\rm P} $ is a countably additive nonnegative
measure on  $\Sigma $
such that  ${\rm P} (\Omega )=1$. An element  $A\in
\Sigma $  is called an \textit{event}, and the measure  ${\rm P} (A)$
is called the \textit{probability of the event\/}  $A$.
Let  $\{ Y,{\rm B}\; {\rm \} }$
be a measurable space. A mapping   $\xi :\Omega \to
Y$ is called {\it $\Sigma |{\rm B}$-measurable\/}, if   $\xi ^{-1}
\left({\rm B}\right)\subset \Sigma $.  A $\Sigma |{\rm B}$-measurable mapping $ \xi$
 is called
\textit{a random variable     with values in  $Y$} and is denoted by $\xi =\xi
\left(\omega \right)$. Such a function  $\xi
\left(\omega \right)$ induces a probability measure  $P_{\xi }
\left(B\right)={\rm P} \left\{\xi ^{-1} \left(B\right)\right\}$
on sets    $B\in {\rm B}$. The function  $P_{\xi }
\left(B\right)$ is called the \textit{distribution function\/} of the
random variable~$\xi $.

A \textit{random mapping\/} of a set   $T$ into  a measurable space  $\{ Y,{\rm B\} }$
is defined as  a mapping   $\xi
\left(t,\omega \right):T\times \Omega \to Y$ such that for any fixed   $t\in T$
it is a measurable mapping from  $\left(\Omega ,\Sigma \right)$ to  $\{ Y,{\rm B\} }$,
i.e., for any   $B\in {\rm B}$, we have
$$
 \left\{\omega \in \Omega :\xi \left(t,\omega \right)\in B\right\}\in \Sigma  .
 $$
If the parameter $t$  is interpreted as   time, a random mapping is called
a \textit{random process}.

Let  $Y\equiv Q_{p} $,  $T\equiv R_{+}^{1} $.  As a probability
space one can take   $\Omega \equiv Q_{p} $ and    $\Sigma  \equiv
{\rm B} $, where $ {\rm B} $ is the   $\sigma $-algebra of all
measurable subsets of   $Q_{p} $. To define a Markov process on $
Q_p$ homogeneous with respect to time, it suffices to define its
distribution density function
 $f\left(x,t\right)$  and the transition density  $f(y,t|x)$
 satisfying the following  conditions:
 \begin{itemize}

\item[1.]  $ f(x,t)$ is ${\rm B}$-measurable in  $x\in Q_{p} $  for any  $t$;

\item[2.]  $\int _{Q_{p} }f\left(x,t\right)d_{p} x =1$;

\item[3.]  $f(y,t|x)\ge 0$  for any  $x\in Q_{p} $,  $y\in Q_{p} $,
and $t>0$;

\item[4.]  $f(y,t|x)$  is  ${\rm  B}\times {\rm B}$-measurable in  $x,\, y$
 for any  $t>0$;

\item[5.]  $\int _{Q_{p} }f(y,t|x)d_{p} y\le 1 $  for any  $x\in
Q_{p} $ and   $t\ge 0$;

\item[6.] for any  $x\in Q_{p} $,  $y\in Q_{p} $,  $s\ge 0$, and  $t\ge 0$,
the Chapman--Kolmogorov condition holds:
\beq
f(z,t+s|x)=\int _{Q_{p} }f(z,t|y)f(y,s|x)d_{p} y;
\label{(1.4)}
\eeq

\item[7.] for any  $x\in Q_{p} $,  $s\ge 0$, and   $t\ge 0$, the compatibility condition holds:
\beq
 f(z,t+s|x)=\int _{Q_{p} }f(z,t|y)f(y,s)d_{p} y  .   \label{(1.5)}
\eeq
\end{itemize}

In this case,   $f\left(x,t\right)$ defines a one-point distribution function for
the random process:
$$
 P\left(B,t\right)=\int _{B}f\left(x,t\right)d_{p} x  ,
 $$
 and
 $f(y,t|x)$  defines the transition function for the homogeneous Markov process:
 $$
 P(B,t|x)=\left\{
\begin{array}{l} {\displaystyle \int _{B}f(y,t|x)d_{p} x,{\rm \; \; }t>0,{\rm \; }x\in Q_{p} {\rm ,\;
\; }B\in {\rm B,\; \; } } \\[10pt]
 {I_{B} (x),{\rm \; \; }t=0,} \end{array}\right.
 $$
 where   $I_{B} (x)$ is the characteristic function of the set $B$.

\section{The First Passage Problem}

Consider a homogeneous Markov process  $\xi
\left(t,\omega \right):R_{+}^{1} \times \Omega \to Q_{p} $ with
the transition density
\beq
 f(y,t|x)\equiv f(y-x,t)=\int _{Q_{p} }\exp (-|k|_{p}^{\alpha } t)
 \chi \left(k(y-x)\right)d_{p} k. \label{(2.1)}
\eeq
The function   $f(y-x,t)$ satisfies the Markovian conditions
\begin{eqnarray*}
&& f(x,t)>0,\quad   \int _{Q_{p} }f(x,t)d_{p} x =1,\\
&&  f(x,t)\to \delta (x) \quad\mbox{in}\quad D' \quad\mbox{as}\quad  t\to 0+\,,\\
&&   \int _{Q_{p} }f(x-y,t)f(y,t') d_{p} y=f(x,t+t') .
\end{eqnarray*}
The function  $f(y-x,t)$ of the form \eqref{(2.1)} is a fundamental solution of
the ultrametric diffusion equation
\beq
 \frac{\partial }{\partial t} f(x,t)=-D_{x}^{\alpha } f\left(x,t\right). \label{(2.2)}
\eeq

This random process was introduced in \cite{VVZ} as a   $p$-adic
analogue of random walk  (on the  $p$-adic line), and equation
\eqref{(2.2)} was interpreted as a   $p$-adic analogue of the
diffusion equation, although the operator  $D_{x}^{\alpha } $ is
nonlocal and its correspondence to the Laplace operator is
problematic. Note that in contrast to Wiener processes, the $
p$-adic random walk $\xi \left(t,\omega \right)$ admits  no
continuous trajectories, since  $Q_{p} $ is a totally disconnected
topological space. The support of  $\xi \left(t,\omega \right)$
belongs to the class of functions without discontinuities of the
second kind (see, for instance, \cite{VVZ}). The operator
$D_{x}^{\alpha } $ can be interpreted in more clear physical terms,
if \eqref{(2.2)} is regarded as a kinetic equation
\cite{ABK}--\cite{ABO}, which is justified  in view of the integral
representation \eqref{(1.3)} of the pseudodifferential operator
$D_{x}^{\alpha } $.

One of the classical problems of random walk on the real line is
that of finding the distribution function of the random variable
describing the first time instant when the wandering particle
returns to the origin. Consider a similar problem for the $ p$-adic
random walk  $\xi \left(t,\omega \right)$ defined above.

Let the evolution of the distribution density function  $\varphi (x,t)$
of the random process  $\xi \left(t,\omega \right)$ be described by the Cauchy problem
for the ultrametric diffusion equation
\beq
\frac{\partial }{\partial t} \varphi (x,t)=-
\frac{1}{\Gamma _{p} (-\alpha )} \int _{Q_{p} }d_{p} y\frac{\varphi (y,t)-
\varphi (x,t)}{\left|y-x\right|_{p}^{\alpha +1} }   ,  \label{(2.3)}
\eeq
with the initial condition
\beq
\varphi (x,0)=\Omega (|x|_{p} ) ,\label{(2.4)}
\eeq
where   $\Omega (|x|_{p} )=\left\{\begin{array}{l} {1,{\rm \; \; \; }|x|_{p} \le 1,}
   \\ {0,{\rm \; \; \; }|x|_{p} >1} \end{array}\right. $
     is the characteristic function of the domain  $Z_{p} $.
\medskip

\noindent{\bf Definition.} The random variable
 $\tau _{Z_{p} } \left(\omega \right)\, \,
   :\Omega \to R_{+}^{1} $
defined by the relation
$$
 \tau _{Z_{p} }
\left(\omega \right)=\inf \left\{\; t>0:\left|\xi \left(t,\omega \right)\right|_{p}
\le 1,\, \, if\, \, \exists \; t':\; \left|\xi \left(t',\omega \right)\right|>1\; ,\;
 0<t'<t\right\}
$$
is called the \textit{first passage time} of a trajectory of the random process
$\xi \left(t,\omega \right)$ entering the domain $ Z_p$ (i.e., the first instant
when it returns to $ Z_p$).
\medskip

The initial condition  \eqref{(2.4)} obviously implies that
$$
{\rm P} \{ \omega \in \Omega :\left|\xi
 \left(0,\omega \right)\right|\le 1\} =1.
 $$
 \medskip

\noindent{\bf Theorem 1. }{\it The distribution density function
$f\left(t\right)$  of the random variable $\tau _{Z_{p} }
\left(\omega \right)$ satisfies the nonhomogeneous Volterra equation
$$
 g(t)=\int _{0}^{t}g(t-\tau )f(\tau ) d\tau +f(t)
 $$
 with}
$$ g(t)=-\frac{1}{\Gamma _{p} (-\alpha )}
 \int _{Q_{p} \backslash Z_{p} }
 \frac{\varphi (x,t)}{\left|x\right|_{_{p} }^{\alpha +1} }  dx.
 $$
\medskip
{\it Proof}. Consider the event  $A\left(t_{i} ,t_{j}
\right)$ that consists in that a particle staying in
the domain
$Q_{p} \backslash Z_{p} $ goes back to the domain   $Z_{p} $
at a time belonging to the interval   $(t_{i} ,t_{j} ]$  (under the condition that
at  $t=0$  the particle stays in   $Z_{p} $):
\begin{eqnarray*}
&&\hskip-1em A\left(t_{i} ,t_{j} \right)=\\
&&\hskip1em = \{ \omega \in \Omega
:\quad \exists t\in (t_{i} ,t_{j} ],\; \mathop{\lim
}\limits_{t'\to t-0} \xi \left(t',\omega \right)\notin Z_{p} ,\;
\mathop{\lim }\limits_{t'\to t+0} \xi \left(t',\omega \right)\in
Z_{p} |_{\; } \xi \left(0,\omega \right)\in Z_{p} \}.
\end{eqnarray*}
Consider also the event  $B\left(t_{i} ,t_{j} \right)$ that consists in that a particle
staying in the domain   $Q_{p}
\backslash Z_{p} $ goes back  to the domain    $Z_{p} $ for the first time at
an instant belonging to the interval $(t_{i} ,t_{j} ]$:
$$
 B\left(t_{i,} t_{j} \right)=\{ \omega \in \Omega :t_{i} <\tau _{Z_{p} }
\left(\omega \right)\le t_{j} \}.
$$
Let us divide the interval   $\left(0,t\, \right]$  into   $n$ parts:
$$0\equiv t_{0} <t_{1} <t_{2} <\ldots<t_{n-1} <t_{n} \equiv t.$$
We obviously have  $A\left(t_{n-1} ,t_{n} \right)\subset \bigcup
_{i=1}^{n}B\left(t_{i-1} ,t_{i} \right) $.
Since
$A\left(t_{n-1} ,t_{n} \right)\bigcap B\left(t_{n-1} ,t_{n}
\right) =B\left(t_{n-1} ,t_{n} \right)$, it follows that
\begin{eqnarray}
A\left(t_{n-1} ,t_{n}
\right)&=&A\left(t_{n-1} ,t_{n} \right)\bigcap \left(\bigcup
_{i=1}^{n}B\left(t_{i-1} ,t_{i} \right) \right) =\nonumber\\
&=&\bigcup
_{i=1}^{n}\left(A\left(t_{n-1} ,t_{n} \right)\bigcap
B\left(t_{i-1} ,t_{i} \right) \right) =\nonumber \\
&=&
\left\{\bigcup
_{i=1}^{n-1}\left(A\left(t_{n-1} ,t_{n} \right)\bigcap
B\left(t_{i-1} ,t_{i} \right) \right) \right\}\bigcup
B\left(t_{n-1} ,t_{n} \right) \label{(2.5)}
\end{eqnarray}

Let   ${\rm P} \left\{A\left(t_{i-1} ,t_{i} \right)\right\}$ and
${\rm P} \left\{B\left(t_{i-1} ,t_{i} \right)\right\}$ be the probabilities of the
events  $A\left(t_{i-1} ,t_{i} \right)$ and  $B\left(t_{i-1} ,t_{i} \right)$,
respectively. Taking into account \eqref{(2.5)} and the incompatibility of the
events  $B\left(t_{i-1} ,t_{i} \right)$, we can write
\begin{eqnarray}
&& \hskip-1em {\rm P} \left\{A\left(t_{n-1} ,t_{n} \right)\right\}= \label{(2.6)}  \\
&&= \sum _{i=1}^{n-1}{\rm P} \left\{A\left(t_{n-1} ,t_{n} \right)
\bigcap B\left(t_{i-1} ,t_{i} \right) \right\}+{\rm P} \left\{B\left(
t_{n-1} ,t_{n} \right)\right\}=\nonumber\\
&&=\sum _{i=1}^{n-1}{\rm P} \left\{A\left(t_{n-1} ,t_{n} \right)
\left|B\left(t_{i-1} ,t_{i} \right)\right. \right\}{\rm P}
\left\{B\left(t_{i-1} ,t_{i} \right)\right\}+{\rm P}
\left\{B\left(t_{n-1} ,t_{n} \right)\right\}= \nonumber\\
&&=
\sum _{i=1}^{n-1}\left\{{\rm P} \left\{A\left(t_{n-1} -t_{i} ,
t_{n} -t_{i} \right)\right\}+
\varepsilon (t_{i} -t_{i-1} )\right\}{\rm P}
\left\{B\left(t_{i-1} ,t_{i} \right)\right\}+{\rm P}
\left\{B\left(t_{n-1} ,t_{n} \right)\right\}  . \nonumber
\end{eqnarray}
Here, we have used the relation
$$
 {\rm P} \left\{A\left(t_{n-1} ,t_{n} \right)
 \left|B\left(t_{i-1} ,t_{i} \right)\right. \right\}=
 {\rm P} \left\{A\left(t_{n-1} -t_{i} ,t_{n} -t_{i} \right)\right\}+
 \varepsilon (t_{i} -t_{i-1} ),
 $$
where  $\varepsilon (t_{i} -t_{i-1} )\to 0$ as  $t_{i} -t_{i-1}
\to 0$.  On the other hand, the probability  ${\rm P}
\left\{A\left(t_{n-1} ,t_{n} \right)\right\}$ is determined by the solution  $\varphi (x,t)$
of the Cauchy problem for the ultrametric diffusion equation
 \eqref{(2.3)} with the initial condition \eqref{(2.4)} and has the form
\beq
{\rm P} \left\{A\left(t_{n-1} ,t_{n}
\right)\right\}=g(t_{n} )\left(t_{n} -t_{n-1}
\right)+o\left(t_{n} -t_{n-1} \right) ,
\label{(2.7)}
\eeq
where  $g(t)$ is defined by
\beq
 g(t)=-\frac{1}{\Gamma _{p} (-\alpha )}
\int _{Q_{p} \backslash Z_{p} }\frac{\varphi
(x,t)}{\left|x\right|_{_{p} }^{\alpha +1} }  dx . \label{(2.8)}
\eeq

The function   $g(t)$ is interpreted as the density of the probability
to go back to the domain $Z_{p} $ at time $t$. Similarly,
the probability  $P\left\{B\left(t_{i-1} ,t_{i} \right)\right\}$
of the first passage to the domain  $Z_{p} $ on the time interval  $\left(t_{i-1} ,t_{i} \right]$
for the same random process can be represented in the form
\beq
 {\rm P} \left\{B\left(t_{i-1} ,t_{i}
\right)\right\}=f(t_{i} )\left(t_{i} -t_{i-1}
\right)+o\left(t_{i} -t_{i-1} \right) ,\label{(2.9)}
\eeq
where    $f(t_{i} )$ is the sought density of the probability of
the first passage to the domain   $Z_{p} $ at time  $t_{i} $.
Now, substituting \eqref{(2.7)} and \eqref{(2.9)} into \eqref{(2.6)}
and passing to the limit as  $\max_{i=1,\ldots,n} \left\{\left(t_{i} -t_{i-1}
\right)\right\}\to 0$, we obtain a nonhomogeneous Volterra equation
of convolution type,
\beq g(t)=\int _{0}^{t}g(t-\tau )f(\tau ) d\tau +f(t).
\label{(2.10)}
\eeq

Note that   $g(t)$ is a continuous function, and therefore, equation
\eqref{(2.10)} has a unique solution in the class of continuous
functions (see, for instance, \cite {VL}). It is easy to check that
$g(t)$ is a function with a finite growth exponent for $t\ge 0$, and
therefore,  $f(t)$ has a finite growth exponent for $t\ge 0$ and
there exist Laplace transforms of the functions $g(t)$, $f(t)$
denoted by $ G(s)$, $ F(s)$, respectively. Passing to the Laplace
transforms in \eqref{(2.10)}, it is easy to find that \beq
F(s)=\frac{G(s)}{1+G(s)}. \label{(2.11)} \eeq Let us calculate
$G(s)$. Substituting the solution of the Cauchy problem
\eqref{(2.3)}--\eqref{(2.4)}, which has the form (see \eqref{(2.1)},
\eqref{(2.2)})
$$
 \varphi (x,t)=\int _{Q_{p} }
\Omega \left(\left|k\right|\right)\exp
\left[-\left|k\right|_{p}^{\alpha } t\right]\chi
\left(-kx\right) d_{p} k,
$$
into \eqref{(2.8)}, integrating the result in  $x$, and then
passing to the Laplace transforms in  $t$, we get
\beq
G(s)=\int _{Q_{p} }\Omega (|k|_{p} )\frac{B_{\alpha }
 -|k|_{p} ^{\alpha } }{s+|k|_{p} ^{\alpha } }  d_{p} k=(B_{\alpha } +s)J(s)-1,
 \label{(2.12)}
 \eeq
where
\begin{eqnarray}
&& B_{\alpha } =\frac{(1-p^{-1} )}{1-p^{-\alpha -1} },
\quad p^{-\alpha } <B_{\alpha } <1 , \nonumber\\
&&  J(s)=\left(1-p^{-1}
\right)\sum _{n=0}^{\infty }p^{-n} \frac{1}{s+p^{-\alpha n} }\;.  \label{(2.13)}
\end{eqnarray}
Substituting \eqref{(2.12)} into \eqref{(2.11)}, we obtain the Laplace transform
of the desired function:
\beq
 F(s)=1-
\frac{1}{\left(B_{\alpha } +s\right)J(s)}\; .\label{(2.14)}
\eeq
The function   $F(s)$ is analytic in the domain  ${\rm Re}\,s>0$ and
tends to zero as $ |s|\to\infty$, uniformly with respect to   $\arg s$.
The function  $F(s)$ is the Laplace transform of the function  $f(t)$
with zero growth exponent:
$\left|f(t)\right|<M$.  Now, it is not difficult to show that $ f(t)$
has the following properties:\smallskip

1. For  $\alpha \ge 1$, we have   $\int _{0}^{\infty }f(t)dt =F(0)=1$, which means
that for   $\alpha \ge 1$ the particle is sure to  return to the initial region,
and therefore,  on an infinite time interval will go back to that region
infinitely many times. In this case, however, there is no finite mean waiting time
 for the first passage:
$$
\left\langle \tau _{Z_{p} } \right\rangle =
\int _{0}^{\infty }tf\left(t\right)dt = -\mathop{\mathop{\lim
}\limits_{s\to 0} } \limits_{{\rm Re\,}s>0} \frac{d}{ds} F(s)\to +\infty
{\kern 1pt} \, .
$$

2. For  $0<\alpha <1$, we have  $\int _{0}^{\infty }f(t)dt
=F(0)=\frac{p}{p^{\alpha } } \left(\frac{p^{\alpha } -1}{p-1}
\right)^{2} \equiv \; \; C_{\alpha } <1$. This means that for small
  $\alpha $, there exist trajectories of the unltrametric random walk
that abandon the initial region never to go back. Note that for the real-valued
Brownian  motion the return property of its trajectories is
missing only if the dimension of the space is greater than two.

Consider more closely the function   $F(s)$.  Clearly,   it has simple poles
at  $s=-\lambda _{k} $,
  $k=0,1,2,\ldots$, which are simple roots of the equation
    $J(s)=0$, and
   $s=-B_{\alpha } \equiv -\lambda _{-1} $. From \eqref{(2.13)}, it is easy to see that
    the values  $\lambda
_{k} $ belong to the interval  $p^{-\alpha (k+1)} <\lambda _{k}
<p^{-\alpha k} $. The point $s=0$  is essentially singular and is a limit point of the poles.
The function  $F(s)$ is non-meromorphic on the complex plane, and this
is an obstacle to the application of the residue theory for the calculation of
the inverse Laplace transforms. To overcome this obstacle, we first prove the following result.
\medskip

\noindent{\bf Lemma 1.} {\it  The function $F(s)$ can be represented as an infinite sum
of terms with simple
poles at the points  $s=-\lambda _{k} $, $ k=-1,0,1,2,\ldots$, namely,
\beq
 F(s)=\sum _{k=-1}^{\infty }\frac{b_{k} }{s+\lambda _{k} }, \label{(2.15)}
\eeq
where $b_{k} $ are the residues of  $F(s)$ at the points $-\lambda _{k} $.
On any closed set $ G $  that does not contain  $ s=0$,
the series \eqref{(2.15)} becomes uniformly convergent, if
 its finitely many terms with poles in $ G$ are dropped. }
\medskip

\noindent{\it Proof.} Consider the auxiliary function
$$
\Phi (w)=F\left(\frac{1}{w} \right) ,\quad \mathop{\lim }\limits_{w\to
0} \Phi (0)=\mathop{\lim }\limits_{s\to \infty } F(s)=0 ,\quad
\mathop{\lim }\limits_{w\to \infty } \Phi (w)=F(0) .
$$
The function  $\Phi (w)$ is analytic   on the complex plane except
at   the simple poles  $w_{-1} =-\frac{1}{B_{\alpha } } $, $w_{k}
=-\frac{1}{\lambda _{k} } $ ,  $k=0,1,2\ldots$. By the
Mittag--Leffler theorem (see, for instance, \cite{LS}),  $\Phi (w)$
can be represented in the form
$$
 \Phi (w)=\sum _{k=-1}^{\infty }\left(\frac{c_{k} }{w-w_{k} } -p_{k} \right) +c ,
$$
where $ c$ is a constant and   $c_{k} $  are the residues of  $\Phi (w)$,
and this series becomes uniformly convergent on any closed bounded set, if its terms
with  poles in that set are dropped. Then
\begin{eqnarray*}
 F(s)&=&\sum _{k=-1}^{\infty }\left(\frac{c_{k}
}{\frac{1}{s} +\frac{1}{\lambda _{k} } } -p_{k} \right) +c=\sum
_{k=-1}^{\infty }\left(\frac{c_{k} \lambda _{k} s}{s+\lambda
_{k} } -p_{k} \right) +c=\\
&=&\sum _{k=-1}^{\infty }\left(\frac{c_{k}
\lambda _{k}^{2} }{s+\lambda _{k} } +c_{k} \lambda _{k} -p_{k}
\right) +c ,
\end{eqnarray*}
 and since  $\Phi (0)=0$, we have
$$
F(s)=\sum _{k=-1}^{\infty }\frac{c_{k} \lambda _{k}^{2} }{s+\lambda _{k} }\;.
$$
Letting  $c_{k} \lambda _{k}^{2} =b_{k} $, we obtain \eqref{(2.15)}.

From Lemma 1 and the uniform convergence
of  \eqref{(2.15)} in the domain
 ${\rm Re}\,s\ge s_{0} >0$, we see that for the calculation of the original
 function  $f(t)$ it suffices to apply the inverse Laplace
 transformation to the series \eqref{(2.15)} term-by-term.
Thus, we get
$$
 f(t)=L_{s\to t}^{-1} [F(s)](t)=\sum _{k=-1}^{\infty }b_{k} L_{s\to t}^{-1}
 \left[\frac{1}{s-\lambda _{k} } \right] (t) ,
$$
and finally,
\beq
f(t)=\sum _{k=-1}^{\infty }b_{k} \exp [-\lambda _{k} t]  ,
\label{(2.16)}
\eeq
where
\begin{eqnarray}
 b_{-1}& =&\frac{1}{J(-B_{\alpha } )}  , \label{(2.17)}\\
 b_{k}& =&\frac{1}{(B_{\alpha } -\lambda _{k} )} \frac{\left(1-p^{-1}
  \right)^{-1} }{\sum _{n=0}^{\infty }\frac{p^{-n} }{(\lambda _{k}
  -p^{-\alpha n} )^{2} }  } ,  \qquad k=0,1,2,\ldots.\label{(2.18)}
\end{eqnarray}

It is not difficult to see that the series  $\sum _{k=0}^{\infty }b_{k}  $
is convergent and  majorizes the series \eqref{(2.16)}, which implies
uniform convergence  of the latter and the continuity of   $f(t)$.

The above results can be summed up as follows:\medskip

\noindent{\bf Theorem  2.} {\it The distribution density for the
first passage times of a trajectory of the ultrametric random walk
can be represented as a uniformly convergent series \eqref{(2.16)}
whose coefficients are defined by \eqref{(2.17)} and
\eqref{(2.18)}.}
\medskip

Let us  go on with the examination of   $f(t)$. It is not difficult to show that
\smallskip

1)  $\mathop{\lim }\limits_{t\to 0} f(t)=\mathop{\lim
}\limits_{s\to \infty } sF(s)=0$;
\smallskip

2)  $\mathop{\lim }\limits_{t\to \infty }
f(t)=\mathop{\mathop{\lim }\limits_{s\to 0} }\limits_{{\rm Re}\,s>0}
sF(s)=0$.\medskip

Then, since the function  $f(t)$ is positive and continuous, it must have a maximum.
Let us show that this maximum is unique.

From the first limit, we have  $\sum _{k=-1}^{\infty }b_{k}  =0$.
Thus, the series \eqref{(2.16)} can be represented as the difference of
two monotonically decreasing strictly concave down functions,
 $f(t)=-f_{1} (t)+f_{2} (t)\ge 0$, and therefore,  $\mathop{\sup
}\limits_{t\in R_{+} } \left|f_{1} (t)-f_{2} (t)\right|$ is unique.

The asymptotic behavior of the function  $f(t)$ for all  $\alpha $
is described by the following theorem.
\medskip

\noindent{\bf Theoremа 3. }{\it For the first passage time
distribution density $ f(t)$ the following estimates hold{\rm :}
\begin{eqnarray}
 A(\alpha )t^{-\frac{2\alpha-1}{\alpha } }
 \left(1+o(1)\right)&\le & f(t)\le B(\alpha )t^{-\frac{2\alpha-1}{\alpha } }
  \left(1+o(1)\right)\; \quad\mbox{\rm for}\quad\alpha >1; \label{(2.19)}\\
A(\alpha )t^{-\frac{1}{\alpha } } \left(1+o(1)\right)&\le& f(t)\le B(\alpha )
t^{-\frac{1}{\alpha } } \left(1+o(1)\right)
\quad\mbox{\rm for}\quad \alpha <1; \label{(2.20)}\\
A(\alpha )\frac{t^{-1} }{(\ln t)^{2} } \left(1+o(1)\right)&\le& f(t)\le B(\alpha )
\frac{t^{-1} }{(\ln t)^{2} } \left(1+o(1)\right)\quad\mbox{\rm for}\quad\alpha =1,
\label{(2.21)}
\end{eqnarray}
where  $o(1)\to 0$ as
$t\to \infty $, and  $A(\alpha )$,  $B(\alpha )$
are functions of  $\alpha $ and  $p$.}
\bigskip

This theorem is proved in Appendix B.

\section{{\Large\bf\it p\,}-Adic Analogue of
the Diffusion Equation with Absorbing Region
for the First Passage Problem}

For the classical problem of random walk of a particle on a straight
line, it is well-known that the distribution density function for
the first instant at which the particle reaches a given domain can
be found from the solution of the diffusion equation with an
absorbing region (see, for instance, \cite{VK}). We are going to
show that a similar approach can be used in the $ p$-adic case: the
distribution density function $f(t)$ for the time  of the first
return to the domain $Z_{p} $ can be obtained from the solution of
the Cauchy problem for the ultrametric diffusion equation with the
absorbing region  $Z_p $, i.e., the  equation \beq
 \frac{\partial \psi (x,t)}{\partial t} =-\frac{1}{\Gamma _{p} (-\alpha )}
 \left(\int _{Q_{p} }\frac{\psi (y,t)-\psi (x,t)}{|x-y|_{p} ^{\alpha +1} }
  d_{p} y-\Omega (|x|_{p} )\int _{Q_{p} \backslash Z_{p} }
  \frac{\psi (y,t)}{|x-y|^{\alpha +1} }  d_{p} y\right),   \label{(3.1)}
  \eeq
with the initial condition  $\psi (x,0)=\Omega (|x|_{p} )$.
The second term in the right-hand side of equation \eqref{(3.1)} is equal to the
probability of transition from the region $Q_{p} \backslash Z_{p} $
to the absorbing region  $Z_{p} $  per unit time. Since this transition for
all trajectories of the random walk \eqref{(3.1)} is always the first one, it follows that
the probability density of this passage at the instant $ t$ is defined by the formula
\beq
 f(t)=-\frac{1}{\Gamma _{p} (-\alpha )} \int _{Q_{p} \backslash Z_{p} }
 \frac{\psi (x,t)}{\left|x\right|_{_{p} }^{\alpha +1} }  dx. \label{(3.2)}
\eeq

Thus, we have two approaches to  finding a solution of the first passage problem.
Their equivalence is established  by the following theorem.
\medskip

\noindent{\bf Теорема 4. }{\it The first passage time distribution density function
obtained from the solution of the Cauchy problem for the ultrametric diffusion equation
with the absorbing region $ Z_p$ coincides with the solution of the Volterra equation
\eqref{(2.10)} }
\medskip

\noindent{\it Proof:} Let us apply the Fourier   transformation  to
 $\psi(x,t)$ with respect to the $ p$-adic variable  $x$  and then the Laplace
 transformation  with respect to the real variable $ t$.  Denote the resulting
 Fourier--Laplace transform by  $\tilde{\Psi }\left(k,s\right)$.
 From \eqref{(3.1)},
taking into account the initial condition
$\psi (x,0)=\Omega (|x|_{p} )$, we obtain the following
 nonhomogeneous Fredholm equation for $\tilde{\Psi }\left(k,s\right)$:
$$
 s\tilde{\Psi }(k,s)=\Omega (|k|_{p} )-|k|_{p} ^{\alpha } \tilde{\Psi }(k,s)-
 \Omega (|k|_{p} )\int _{Q_{p} }\tilde{\Psi }(q,s)\left(B_{\alpha }
 -|q|_{p} ^{\alpha } \right) \Omega (|q|_{p} )d_{p} q ,
 $$
or
\beq
\tilde{\Psi }(k,s)=\frac{\Omega (|k|_{p} )}{s+\left|k\right|_{p}^{\alpha } }
 -\frac{\Omega (|k|_{p} )}{s+\left|k\right|_{p}^{\alpha } }
 \int _{Q_{p} }\tilde{\Psi }(q,s)\left(B_{\alpha } -|q|_{p} ^{\alpha } \right)
 \Omega (|q|_{p} )d_{p} q . \label{(3.3)}
\eeq
Multiplying equation \eqref{(3.3)} by  $\left(B_{\alpha }
-\left|k\right|_{p}^{\alpha } \right)\Omega
\left(\left|k\right|_{p} \right)$  and integrating the result, we get
\begin{eqnarray}
&&\hskip-1em\int _{Q_{p} }\tilde{\Psi }(k,s)\left(B_{\alpha } -|k|_{p} ^{\alpha } \right)
 \Omega (|k|_{p} )d_{p} k=\int _{Q_{p} }\Omega (|k|_{p} )\frac{B_{\alpha }
  -|k|_{p} ^{\alpha } }{s+|k|_{p} ^{\alpha } }  d_{p} k+\nonumber \\
&&\hskip2em +\int _{Q_{p} }\Omega (|k|_{p} )\frac{B_{\alpha } -|k|_{p} ^{\alpha } }{s+
|k|_{p} ^{\alpha } }  d_{p} k\int _{Q_{p} }\tilde{\Psi }(q,s)
\left(B_{\alpha } -|q|_{p} ^{\alpha } \right) \Omega (|q|_{p} )d_{p} q.\label{(3.4)}
\end{eqnarray}
Note that  $\int _{Q_{p} }\tilde{\Psi }(q,s)\left(B_{\alpha } -|q|_{p} ^{\alpha } \right)
\Omega (|q|_{p} )d_{p} q\equiv F(s)$
is the Laplace transform
of the first passage time distribution density function
$f(t)$  defined by \eqref{(3.2)}. Now, in view of \eqref{(2.12)},
we can rewrite equation \eqref{(3.4)}
in the form
 $$
F(s)=G(s)+G(s)F(s) .
$$
Comparing this with \eqref{(2.11)}, we see that the solution of the last equation
coincides with that of the Volterra equation \eqref{(2.10)}.

\section{Number of Returns   for Ultrametric Diffusion}

In this section, we consider some questions pertaining to the probability
of the $ m$-th return on the time interval $ (0,t]$ and the growth of the number of returns
with the growth of $ t$.

For the probability space   $\{ \Omega ,\Sigma ,{\rm P} \}$,
consider a random process
 $$N_{Z_{p} } (t,\omega
):\Omega \times R_{+}^{1} \to N\subset Z_+,\qquad N_{Z_{p} } (0,\omega )=0
$$
that describes the number of returns of a particle to the domain  $Z_{p} $
on a finite time interval  $(0,t]$.  Let us calculate the probability
of the   $m$-th return of a particle to   $Z_{p} $  on the interval
$(0,t]$. Consider the event   $Q_{t}^{m} =\{ \omega \in \Omega
:\; N_{Z_{p} } (t,\omega )\ge m\} $ that consists in that a particle staying in the domain
 $Q_{p} \backslash Z_{p} $ goes back to    $Z_{p} $
 for the $ m$-th time at an instant from the interval  $(0,t]$,
 or equivalently, that  a particle  visits  the domain $ Z_p$ at least $ m$ times
 on the time  interval   $(0,t]$. Denote the probability of this event by
 ${\rm P} \{ Q_{t}^{m} \} =q^{(m)} (t)$. Obviously, $ Q_{0}^{m} =\emptyset$ for
 all $ m>0$ and  $ Q_t^0=\Omega$ for all $ t>0$.
\medskip

\noindent{\bf Теорема 5.} {\it  The probability  $q^{(m)} (t)$  of the $m$-th return on
the interval
  $(0,t]$ satisfies the recurrent equation
  \beq
   \begin{array}{l} \displaystyle {q^{(m)} (t)=\int _{0}^{t}q^{(m-1)}
(t-\tau )f(\tau )d\tau  \, ,\quad m\ge 1\, ,} \\[12pt]
{q^{(0)}
(t)=1,\quad m=0}, \end{array}
\label{(4.1)}
\eeq
where
$f(t)$  is the distribution density for the
first return time. }
\medskip

\noindent{\it Proof. } This statement is proved along the same lines as
Theorem 1, and therefore, we just outline the main steps.

Consider the event $Q_{t}^{m} $. Let  $B_{\tau +d\tau } =\{
\omega \in \Omega :\, \tau <\tau _{Z_{p} } (\omega )\le \tau
+d\tau \} $ be the event of the first return to
the domain  $Z_{p} $
on the time interval  $(\tau ,\; \tau +d\tau ]$.  Then,
$Q_{t}^{m} \subset \bigcup _{\tau \in (0,t]}B_{\tau +d\tau }  $
and for  $m\ne 0$  we can write
$$ Q_{t}^{m} =Q_{t}^{m} \bigcap \left(\bigcup _{\tau \in (o,t]}B_{\tau +d\tau }
 \right) =\bigcup _{\tau \in (0,t]}\left(Q_{t}^{m} \bigcap B_{\tau +d\tau }
 \right).
 $$
Since the events  $B_{\tau +d\tau } $ are incompatible for all  $\tau \in (0,t]$,
we have
$$
{\rm P} \{ Q_{t}^{m} \} =\sum _{\tau \in (0,t]}{\rm P} \{ Q_{t}^{m}
\bigcap B_{\tau +d\tau }  \}  =
\sum _{\tau \in (0,t]}{\rm P} \{ B_{\tau +d\tau } \}
 {\rm P} \{ Q_{t}^{m} |B_{\tau +d\tau } \}.
  $$
Observing that  ${\rm P} \{ Q_{t}^{m} \left|B_{\tau +d\tau }
\right. \} ={\rm P} \{ Q_{t-\tau }^{m-1} \} $, we obtain
\beq
 \begin{array}{ccl} {\rm P} \{ Q_{t}^{m} \} &=&\displaystyle\sum _{\tau \in (0,t]}{\rm P}
  \{ B_{\tau +d\tau } \}  {\rm P} \{ Q_{t-\tau }^{m-1} \} ,\quad m\ge 1, \\[15pt]
   {\rm P} \{ Q_{t}^{0} \}& =&1\, ,\quad m=0. \end{array}\label{(4.2)}
 \eeq
Finally, recalling that   ${\rm P} \{ Q_{t}^{m} \}
=q^{(m)} (t)$ and using the symbolic formula   ${\rm P} \{ B_{\tau
+d\tau } \} =f(\tau )d\tau $ (its meaning is clear from the rigorous
arguments  in the proof of Theorem 1), we obtain the desired recurrent relation
\eqref{(4.1)}.\medskip

For the Laplace transforms, equation \eqref{(4.1)} reads
\beq \begin{array}{l} {Q^{(m)} (s)=Q^{(m-1)} (s)F(s)}, \\[5pt]
 {Q^{(0)} (s)=s^{-1}, }
 \end{array}        \label{(4.3)}
 \eeq
where  $Q^{(m)} (s)$ is the Laplace transform of  $q^{(m)} (t)$.

The solution of \eqref{(4.3)} has the form
\beq
 Q^{m} (s)=\frac{1}{s} \left(F(s)\right)^{m},
 \label{(4.4)}
 \eeq
where  $F(s)$ is defined by \eqref{(2.14)}.

Using the recurrent equation \eqref{(4.1)}
and the properties of
$f\left(t\right)$  and  $F(s)$, it is not difficult to show
that the functions $q^{(m)} (t)$ have the following properties:
\medskip

 1.    Each  $q^{(m)} (t)$ is a monotonically increasing function of $ t$;
 the function  $\frac{d}{dt} q^{(m)} (t)\ge 0$
 has the meaning of probability density for the $ m$-th return.\medskip

2.     $\displaystyle\lim_{t\to0} \frac{d}{dt} q^{(m)} (t)=0,\quad
\lim_{t\to\infty} \frac{d}{dt} q^{(m)} (t)=0 $.\vskip7pt

3.     $\mathop{\lim }\limits_{t\to 0} q^{(m)} (t)=0$.

4.     $\; \mathop{\lim }\limits_{t\to \infty } q^{(m)}
(t)=\left\{\begin{array}{cl} 1, & \alpha \ge 1, \\
(C_{\alpha } )^{m} ,& \alpha <1.  \end{array}\right.  $
Recall that the quantity  $C_{\alpha } =\frac{p}{p^{\alpha } }
\left(\frac{p^{\alpha } -1}{p-1} \right)^{2} <1$  (see Section 2, property 2 of  $f(t)$)
is the measure of return trajectories of the ultrametric random walk, which is the same
as the probability of the first return on the infinite time interval.

 5.   $q^{(m)} (t)<q^{(m-1)} (t)$, $m\ge 1$, i.e., the sequence  $q^{(m)} (t)$
 is monotonically increasing with respect to
 $ m$ for any  $ t$.
 \bigskip

With the help of the recurrent equation
 \eqref{(4.1)} and  properties 1, 2 of the function
$q^{(m)} (t)$, it is not difficult to show that
the return probability densities have a maximum, which is unique, and thus,
we have a single-mode distribution.

In view of property 4, it is only for  $\alpha \ge 1$  that one can speak
 about the mean waiting time  of the $ m$-th return.
For $\alpha \ge 1$, the $ m$-th (in particular, the first)
return  is a certain event,   but its  mean waiting time  is infinite.
Indeed,
$$
 \int _{0}^{\infty }t\frac{d}{dt} q^{(m)} (t)dt =
\lim_{\stackrel{s\to 0}{{\rm Re}\,s>0}} \left(-m\left(F(s)\right)^{m-1} \frac{d}{ds}
F(s)\right)=\infty,
$$
since  $\lim\limits_{s\to 0,\, {\rm Re}\,s>0} \frac{d}{ds} F(s)=-\infty $.

Next, consider the problem of finding the probability
of precisely   $m$
returns on the time interval   $(0,t]$.
Let  $H_{t}^{m} =\{
\omega \in \Omega :\; N_{Z_{p} } (t,\omega )=m\} $  be the event that
on the time interval $ (0,t]$, the
 particle goes back to the region $ Z_p$  precisely $ m$-times.
We are interested in the probability of this event, ${\rm P} \{ H_{t}^{m} \} =h^{(m)} (t)$.
\medskip

\noindent{\bf Теорема 6.} {\it  The probability $h^{(m)} (t)$ of precisely $ m$ returns
on the time interval $ (0,t]$
satisfies the following recurrent equations{\rm \,:}
\beq
\begin{array}{rcl} h^{(m)} (t)&=&\displaystyle \int _{0}^{t}h^{(m-1)} (t-\tau )f(\tau )d
\tau  \, ,\quad m\ge 1, \\[15pt]
h^{(0)} (t)&=&1-\displaystyle\int _{0}^{t}f(\tau
) d\tau  \, ,\quad m=0\, .
 \end{array} \label{(4.5)}
\eeq
where   $f(t)$ is the probability density for the first return time.
}\medskip

\noindent{\it Proof}: We obviously have  $H_{t}^{m} =Q_{t}^{m}
\backslash {\rm Q}_{t}^{m+1} $, and therefore,
\beq
 h^{(m)} (t)=q^{(m)} (t)-q^{(m+1)} (t). \label{(4.6)}
\eeq
Substituting \eqref{(4.1)} into \eqref{(4.6)}, we obtain the recurrent
equations \eqref{(4.5)}. The theorem is proved.
\medskip

Let us examine more closely the   probability distribution function for
precisely $ m$ returns. In terms of Laplace transforms, equation \eqref{(4.5)}
has the form
\beq
\begin{array}{rcl}
H^{(m)} (s)&=&H^{(m-1)} (s)F(s), \\[5pt]
 H^{(0)} (s)&=&s^{-1}
\left(1-F(s)\right),
\end{array} \label{(4.7)}
\eeq
where  $H^{(m)} (s)$  is the Laplace transform of   $h^{(m)} (t)$.
From \eqref{(4.7)}, we obtain the following expression for
the Laplace transform of the solution of equation
\eqref{(4.5)}:
\beq H^{(m)} (s)=\frac{1}{s}
 \left(1-F(s)\right)\left(F(s)\right)^{m}  .\label{(4.8)}
 \eeq
Using \eqref{(2.14)} and the solution \eqref{(4.8)}, it is not difficult
to establish the following properties of
$h^{(m)} (t)$:
\begin{itemize}

\item[1.]    $h^{(m)} (t)$  is a positive  function such that
 $$\mathop{\mathop{\lim }\limits_{t\to 0}
}\limits_{t\to \infty } h^{(m)} (t)=\mathop{\mathop{\lim
}\limits_{s\to \infty } }\limits_{s\to 0}
(1-F(s))\left(F(s)\right)^{m} =0,\quad \alpha \ge 1 ;
$$

\item[2.]     $\mathop{\lim }\limits_{t\to 0} h^{(m)} (t)=0,\quad
\alpha <1$,\quad  $\displaystyle \mathop{\lim }\limits_{t\to \infty }
h^{(m)} (t)=(1-C_{\alpha } )\left(C_{\alpha } \right)^{m} \;
,\quad \alpha <1$ ;

\item[3.]     $h^{(m)} (t)$  has a maximum, which is unique;

\item[4.]     $h^{(m)} (t)<h^{(m-1)} (t)$.

\end{itemize}

What is the mean number of returns  $\mu (t)$ on the time interval   $(0,t]$?
Usually, it is expected that the mean number of returns is proportional to the
walk  time. By definition, we have
\beq
 \mu (t)=\sum _{n=1}^{\infty }nh^{(n)} (t). \label{(4.9)}
 \eeq
 \medskip

\noindent{\bf Theorem 7. }{ \it The mean number of returns on the
time interval  $(0,t]$ is determined by the formula
\beq
 \mu (t)=\int _{0}^{t}g(\tau )d\tau,
 \label{(4.10)}
 \eeq
where $g(t)$  is defined by \eqref{(2.8)} and is
 the density of the probability to return to the domain
 $Z_{p}$ at the instant  $t$.}
 \medskip

\noindent{\it Proof.}  Writing the expression \eqref{(4.9)}
for Laplace transforms and using \eqref{(4.8)}, we obtain
\beq
 {\rm M} (s)=\sum _{n=1}^{\infty }nH^{(n)}
(s)= \frac{1}{s} \left(1-F(s)\right)\sum _{n=1}^{\infty
}n\left(F(s)\right) ^{n}.
\label{(4.12)}
\eeq
Since  $\left|F(s)\right|<1$  for   ${\rm Re}\,s>0$, the series in \eqref{(4.12)}
can be summed and we have
\beq
{\rm M} (s)=\frac{1}{s} \frac{F(s)}{1-F(s)}\;  .\label{(4.13)}
\eeq
Hence, using \eqref{(2.11)}, we get
\beq
 {\rm M} (s)=\frac{1}{s} G(s)  .
 \label{(4.14)}
 \eeq
Applying the inverse Laplace transformation,
we obtain \eqref{(4.10)}.

Now,   let us calculate the average number of
returns on the time interval   $(0,t]$, using \eqref{(4.10)}. Integrating equation
 \eqref{(2.3)}  over the domain  $Z_{p} $
 and taking into account \eqref{(2.8)}, we obtain the following
 equation:
 \beq
  \frac{\partial }{\partial t} S_{Z_{p} } (t)=-B_{\alpha } S_{Z_{p} } (t)+g(t),
  \label{(4.15)}
  \eeq
where  $S_{Z_{p} } (t)=\int _{Z_{p} }\varphi (x,t)dx $. From the solution of the
Cauchy problem for equation \eqref{(2.3)} and the initial condition
\eqref{(2.4)}, we have the following expression for
$S_{Z_{p} } (t)$:
\beq
 S_{Z_{p} } (t)=\left(1-\frac{1}{p}
\right)\sum _{n=0}^{\infty }p^{-n}  \exp \left[-p^{-\alpha n}
t\right]. \label{(4.16)}
\eeq
The series \eqref{(4.16)} is uniformly convergent,  and therefore,
using \eqref{(4.15)}, \eqref{(4.16)} in \eqref{(4.10)}, we easily obtain
\begin{eqnarray}
&&\hskip-2em \mu (t)=B_{\alpha } \left(1-\frac{1}{p}
\right)\sum _{n=0}^{\infty }p^{\left(\alpha -1\right)n}
\left(1-\exp \left[-p^{-\alpha n} t\right]\right)+\nonumber\\
&&\hskip9em
+\left(1-\frac{1}{p} \right)\sum _{n=0}^{\infty }p^{-n} \exp
\left[-p^{-\alpha n} t\right] -1.\label{(4.17)}
\end{eqnarray}

Note that the first series in \eqref{(4.17)}
is convergent for all  $\alpha >0$, although there is no uniform convergence
for  $\alpha \ge 1$. The asymptotic behavior
of the second series is characterized by the function  $t^{-\frac{1}{\alpha } } $
 (see formula \eqref{(A.2)}  in Apendix A).
 \medskip

\noindent{\bf Theorem 8.}  {\it The following asymptotic estimates hold for
the function $\mu (t)$\textup{:}
\begin{eqnarray}
&& \hskip-2em p^{-1} \left(1-\frac{1}{p} \right)\Gamma \left(\frac{1}{\alpha }
\right)\frac{B_{\alpha } }{(\alpha -1)\ln p} t^{\frac{\alpha
-1}{\alpha } } +O(t^{-\frac{1}{\alpha } } )\le \nonumber\\
&&\hskip2em  \le \mu
(t)\, \; \le p\left(1-\frac{1}{p} \right)\Gamma
\left(\frac{1}{\alpha } \right)\frac{B_{\alpha } }{(\alpha
-1)\ln p} t^{\frac{\alpha -1}{\alpha } } +O(t^{-\frac{1}{\alpha
} } ),\quad \alpha >1, \label{(4.18)}\\[10pt]
&&\hskip-2em
 p^{-1} \left(1-\frac{1}{p} \right)\frac{B_{1} }
{\ln p} \Gamma (1)\ln t+O(t^{-1} )\, \; \le \; \mu (t)\; \le\nonumber \\
&&\hskip8em \le  p\left(1-\frac{1}{p}
 \right)\frac{B_{1} }{\ln p} \Gamma (1)\ln t+O(t^{-1} ),\quad \quad
 \alpha =1,\label{(4.19)}\\[10pt]
&&\hskip-2em
\frac{C_{\alpha } }{1-C_{\alpha } } -p^{-(1-\alpha )}
\left(1-\frac{1}{p} \right)\Gamma \left(\frac{1-\alpha }{\alpha }
\right)\frac{B_{\alpha } }{\alpha \ln p} t^{-\frac{1-\alpha }{\alpha }
} (1+O(t^{-1} ))\le \mu (t)\le \nonumber  \\
&&\hskip2em\le  \frac{C_{\alpha }
}{1-C_{\alpha } } -p^{1-\alpha } \left(1-\frac{1}{p} \right)
\Gamma \left(\frac{1-\alpha }{\alpha } \right)\frac{B_{\alpha } }{\alpha
\ln p} t^{- \frac{1-\alpha }{\alpha } } (1+O(t^{-1} )),  \label{(4.20)}\\
&&\hskip25em 0<\alpha <1,\nonumber
\end{eqnarray}
where $\Gamma (x)$ is the gamma-function and  $C_{\alpha }
=\frac{p}{p^{\alpha } } \left(\frac{p^{\alpha } -1}{p-1}
\right)^{2} $  is the probability of the first return on the infinite time interval.}
\medskip

To obtain the asymptotic estimates of   $\mu (t)$ for    $\alpha
\ge 1$, it suffices to integrate the asymptotic
estimates
$$
p^{-1} (1-p^{-1} )B_{\alpha }
\frac{\Gamma (\alpha ^{-1} )}{\alpha \ln p} t^{-\frac{1}{\alpha
} } \le g(t)\le p(1-p^{-1} )B_{\alpha } \frac{\Gamma (\alpha
^{-1} )}{\alpha \ln p} t^{-\frac{1}{\alpha } }
 $$
obtained from an asymptotic estimate for the series   $S(t)$ (see
formula \eqref{(A.2)} in Appendix~A). This integration is justified,
since the function  $g(t)$ and its asymptotic bounds
continuously depend on   $t\ge a$,
$\alpha \ge 1$, are strictly positive for large  $t$, and
 $$
 \int _{a}^{\infty }t^{-\frac{1}{\alpha } }  dt=+\infty.
  $$

To obtain the asymptotic estimates of
   $\mu (t)$  for   $0<\alpha
<1$, it suffices to rewrite \eqref{(4.17)} in the form
\begin{eqnarray*}
 \mu (t)&=&B_{\alpha }
\left(1-\frac{1}{p} \right)\sum _{n=0}^{\infty
}p^{-\left(1-\alpha \right)n} \left(1-\exp \left[-p^{-\alpha n}
t\right]\right) +\\
&&\hskip11em+ \left(1-\frac{1}{p} \right)\sum _{n=0}^{\infty
}p^{-n} \exp \left[-p^{-\alpha n} t\right] -1=  \\
&=&B_{\alpha }
\left(1-\frac{1}{p} \right)\frac{1}{1-p^{-(1-\alpha )} }
-1-B_{\alpha } \left(1-\frac{1}{p} \right)\sum _{n=0}^{\infty
}p^{-\left(1-\alpha \right)n} \exp \left[-p^{-\alpha n} t\right]+\\
&&\hskip11em
+\left(1-\frac{1}{p} \right) \sum _{n=0}^{\infty }p^{-n} \exp
\left[-p^{-\alpha n} t\right]
\end{eqnarray*}
and use formula \eqref{(A.2)} from Appendix A.

\appendix

\section*{Appendix A}
\setcounter{section}{1}
\setcounter{equation}{0}

Here, we obtain an asymptotic estimate for the series
\beq
S(t)=\mathop{\sum }\limits_{i=0}^{\infty } \frac{1}{i^{k} } a^{-i} e^{-(b)^{-i} t},
\quad   t\ge 0 ,\; k\in N ,\;  a>1 ,\;b>1,  \label{(A.1)}
\eeq
for $ t\gg 1$.\medskip

\noindent{\bf Lemma A1.}  {\it For the series \eqref{(A.1)}  the following estimate holds for
 $t\gg1 :$
 \begin{eqnarray}
 &&\left(\ln b\right)^{k-1}
\left(\ln (bt)\right)^{-k} \left(bt\right)^{- \frac{\ln a}{\ln
b} } \Gamma \left(\frac{\ln a}{\ln b} \right) \left(1+o(t)\right)\le
S(t)\le \nonumber\\
&&\hskip4em \le   a\left(\ln b\right)^{k-1} \left(\ln (t)\right)^{-k}
\left(t\right)^{-\frac{\ln a}{\ln b} } \Gamma \left(\frac{\ln a}{\ln
b} \right)\left(1+o(t)\right) , \label{(A.2)}
\end{eqnarray}
where   $\Gamma \left(z\right)$ is the gamma-function.}
\medskip

\noindent{\it Proof.}  Note that    $\frac{1}{x^{k} }
a^{-x} $ is a decreasing function
and  $e^{-(b)^{-x} t} $
is an increasing function of $ x$.    Therefore, on the interval $i\le x\le i+1$
we have the inequality
\beq
 \frac{1}{x^{k} } a^{-x} e^{-b^{-(x-1)} t}
\le a^{-i} e^{-b^{-i} t} \le \frac{1}{(x-1)^{k} } a^{-(x-1)}
e^{-b^{-x} t}  .\label{(A.3)}
\eeq
Integrating \eqref{(A.3)} in   $x$  from  $i$ to  $i+1$, we get
\beq
 a^{-1}
\int _{i}^{i+1}\frac{1}{x^{k} } a^{-(x-1)} e^{-b^{-(x-1)} t} dx
\le a^{-i} e^{-b^{-i} t} \le a\int _{i}^{i+1}\frac{1}{(x-1)^{k}
} a^{-x} e^{-b^{-x} t} dx  . \label{(A.4)}
\eeq
Now, summing the inequalities \eqref{(A.4)} with respect to $ i$ from $ 0$ to $ \infty$,
we find that
\begin{eqnarray*}
 && S_{\min } (t)=a^{-1}
\int _{0}^{\infty }\frac{1}{x^{k} } a^{-(x-1)} e^{-b^{-(x-1)} t}
dx \le  S(t)\le\\
&&\hskip9em \le  a\int _{0}^{\infty }\frac{1}{(x-1)^{k} } a^{-x}
e^{-b^{-x} t} dx =S_{\max } (t),
\end{eqnarray*}
where
\begin{eqnarray*}
&& S_{\min } (t)=\left(\ln b\right)^{k-1}
\left(\ln (bt)\right)^{-k} \left(bt\right)^{-\frac{\ln a}{\ln b}
} \int _{0}^{bt}\left(1-\frac{\ln y}{\ln (bt)} \right)^{-k}
y^{\frac{\ln a}{\ln b} -1} e^{-y}  dy,\\
&&S_{\max } (t)=a\left(\ln b\right)^{k-1} \left(\ln (t)
\right)^{-k} \left(t\right)^{-\frac{\ln a}{\ln b} } \int
_{0}^{t}\left(1-\frac{\ln y}{\ln (t)} \right)^{-k} y^{\frac{\ln
a}{\ln b} -1} e^{-y}  dy.
 \end{eqnarray*}
Let
$$\int _{0}^{x}\left(1-\frac{\ln y}{\ln x}
\right)^{k} y^{z-1} e^{-y} dy =\gamma ^{(k)} (z,x)$$
 and note that
 $\mathop{\lim }\limits_{x\to \infty } \gamma ^{(k)}
\left(z,x\right)=\int _{0}^{\infty }t^{z-1} e^{-y} dy =\Gamma
(z)$ and  $\gamma ^{(0)} \left(z,x\right)=\int _{0}^{x}t^{z-1}
e^{-y} dy =\gamma (z,x)$ . Then, for   $x\gg 1$, we can write
\begin{eqnarray*}
&& S_{\min } (t)=\left(\ln b\right)^{k-1}
\left(\ln (bt)\right)^{-k} \left(bt\right)^{-\frac{\ln a}{\ln b}
} \Gamma \left(\frac{\ln a}{\ln b} \right)\left(1+o(1)\right),\\
&& S_{\max } (t)=a\left(\ln b\right)^{k-1}
\left(\ln (t)\right)^{-k} \left(t\right)^{-\frac{\ln a}{\ln b} }
\Gamma \left(\frac{\ln a}{\ln b} \right)\left(1+o(1)\right) ,
\end{eqnarray*}
and therefore, the estimate \eqref{(A.2)} holds.

\section*{Appendix B}
\setcounter{section}{2}
\setcounter{equation}{0}

\subsection*{Proof of Theorem 3 from Section 2}

To estimate the function  $f(t)$, we first estimate the coefficients
  $b_{k} $ of the series \eqref{(2.16)}. These coefficients coincide with the
  residues of the function  $F(s)$ at the poles   $s=-\lambda
_{k} $,  $k=-1,0,1,2,\ldots $
(see \eqref{(2.17)}, \eqref{(2.18)}):
\begin{eqnarray*}
 b_{-1}& =& -\frac{1}{J(-B_{\alpha } )}  ,\\
 b_{k} &= & -\frac{1}{\left(B_{\alpha } -
\lambda _{k} \right)} \mathop{\lim }\limits_{s\to -\lambda _{k}
} \frac{s+\lambda _{k} }{J(s)} =\frac{1}{\left(B_{\alpha } -
\lambda _{k} \right)} \frac{1}{J'(-\lambda _{k} )} =
\frac{1}{\left(B_{\alpha } -\lambda _{k} \right)} u_{k}  ,
\end{eqnarray*}
where
\beq
u_{k}
=\left(1-p^{-1} \right)^{-1} \left[ \sum _{n=0}^{\infty }
\frac{p^{-n} }{(\lambda _{k} -p^{-\alpha n} )^{2} }  \right]^{-1}  . \label{(B.1)}
\eeq
Recall (see Section 2) that the poles  $s=-\lambda _{k} $,
$k=0,1,2,\ldots$ coincide with the simple roots of the equation
 $\sum _{n=0}^{\infty }p^{-n}
\frac{1}{s+p^{-\alpha n} }  =0$  and the values  $\lambda _{k} $
lie on the interval   $p^{-\alpha (k+1)} <\lambda _{k} <p^{-\alpha
k} $. The point   $s=0$ is a limit point for the set of  poles.
Let us examine the behavior of the poles  $\lambda _{k} $
and the residues   $u_{k} $ for large  $k$.
We  pass from   $\lambda _{k} $ to new variables
$\delta _{k} $, setting
\beq
\lambda _{k} =p^{-\alpha k} \left(p^{-\alpha } +
(1-p^{-\alpha } )\delta _{k} \right)=p^{-\alpha (k+1)}
+p^{-\alpha k} (1-p^{-\alpha } )\delta _{k}  ,\quad  0\le \delta
_{k} \le 1, \label{(B.2)}
\eeq
and let
\beq \nu_{k} =\left(1-p^{-1}
\right)u_{k} \equiv \left[\sum _{n=0}^{\infty }\frac{p^{-n}
}{(p^{- \alpha (k+1)} +p^{-\alpha k} (1-p^{-\alpha } )\delta
_{k} -p^{-\alpha n} )^{2} }  \right]^{-1}  .
\label{(B.3)}
\eeq
It can be shown that \eqref{(B.3)} implies the following inequalities for  $\nu
_{k} $:
\begin{eqnarray}
&& \hskip-3em\frac{1}{p^{(2\alpha -1)k} }
\left[\frac{1-p^{(2\alpha -1)} }{\left(p^{-\alpha } +
(1-p^{-\alpha } )\delta _{k} \right)^{2} \left(1-p^{2\alpha -1}
\right)}
+ \frac{p}{(1-p^{-\alpha } )^{2} \delta _{k} ^{2} }\right.+\label{(B.4)} \\
&&\hskip-2em + \left.\frac{p^{-2} }{\left(1-p^{-\alpha } \right)^{2}
\left(p^{-\alpha } +
  (1-p^{-\alpha } )\delta _{k} \right)^{2}
\left(1-p^{-1} \right)} \right]^{-1} <  \nu _{k}
<\frac{1}{p^{(2\alpha -1)k} } p(1-p^{-\alpha } )^{2} \delta
_{k}^{2}. \nonumber
\end{eqnarray}
Since $\lambda _{k} $,   $k=0,1,\ldots$, are zeroes of the function
 $\sum _{n=0}^{\infty }\frac{p^{-n} }{s-p^{-\alpha n} }$,
 we have
 $$
  \sum _{n=0}^{\infty }
\frac{p^{-n} }{p^{-\alpha (k+1)} +p^{-\alpha k} (1-p^{-\alpha }
)\delta _{k} - p^{-\alpha n} }  =0 ,
$$
which implies the following estimate for $\delta _{k} $:
\begin{eqnarray}
&&\hskip-4em \frac{1}{a_{k} }
\frac{p+1+p^{2-\alpha } a_{k} }{2p^{2-\alpha } }
 \left[1-\left(1-\frac{4p^{2-\alpha } a_{k} }{\left(p+1+p^{2-\alpha }
 a_{k} \right)^{2} } \right)^{1/2 } \right]< \nonumber\\
 &&\hskip15em<\delta _{k}
 <\frac{1}{a_{k} } \frac{1-p^{-\alpha -1} }{(p-1)(1-p^{-\alpha } )}, \label{(B.5)}
\end{eqnarray}
where $ a_k=k$ for $ \alpha=1$ and $ a_k= \frac{1-p^{(1-\alpha
)(k+1)} }{1-p^{1-\alpha } } $ for $ \alpha\neq 1$.  The quantities $ a_k$ have the
following asymptotic behavior for $ k\to\infty$:
\begin{eqnarray*}
&& a_{k} =\frac{1}{1-p^{1-\alpha } } +o(1) \;\;\mbox{for}\;\; \alpha >1,\\
&&
a_{k} =\frac{p^{1-\alpha } }{p^{1-\alpha } -1} p^{(1-\alpha )k}
   \left(1+o(1)\right) \;\;\mbox{for}\;\;  \alpha <1,\\
   && a_k=k\;\;\mbox{for}\;\;\alpha=1, \quad \mbox{where}\;\; o(1)\to 0\;\;\mbox{as}\;\; k\to\infty.
\end{eqnarray*}
Using these relations and \eqref{(B.4)}, \eqref{(B.5)}, it is not difficult to obtain the
 estimates
\begin{eqnarray}
&& D(\alpha )\left(1+o(1)\right)
<\delta _{k} <U(\alpha )\left(1+o(1)\right) ,\nonumber\\
&&\tilde{D}(\alpha )p^{(1-2\alpha )k} \left(1+o(1)\right)<u_{k}
 <\tilde{U}(\alpha )p^{(1-2\alpha )k} \left(1+o(1)\right)\;\;
 \mbox{for}\;\; \alpha >1 ,   \label{(B.6)}\\[10pt]
&& D(\alpha )p^{(\alpha -1)k}
\left(1+o(1)\right)<\delta _{k} <U(\alpha )p^{(\alpha -1)k}
\left(1+o(1)\right),\nonumber\\
&& \tilde{D}(\alpha )p^{-k} \left(1+o(1)\right)<u_{k}
<\tilde{U}(\alpha )p^{-k} \left(1+o(1)\right)\;\;\mbox{for}\;\;\alpha
<1 ,    \label{(B.7)}\\[10pt]
&&  D(\alpha )k^{-1} \left(1+o(1)\right)<\delta _{k}
<U(\alpha )k^{-1} \left(1+o(1)\right),\nonumber\\
&& \tilde{D}(\alpha )p^{-k} k^{-2}
 \left(1+o(1)\right)<u_{k} <\tilde{U}(\alpha ) p^{-k} k^{-2}
 \left(1+o(1)\right)\;\;\mbox{for}\;\; \alpha =1 ,    \label{(B.8)}
 \end{eqnarray}
where  $D(\alpha )$,   $U(\alpha )$,  $\tilde{D}(\alpha )$,   $\tilde{U}(\alpha )$
are functions of  $\alpha $ and  $p$ whose expressions are too lengthy to be written
out here.

Let $ f(t)$ be  the probability density function \eqref{(2.16)} for the first passage times.
Taking into account the above notation, we can write
\begin{eqnarray*}
&& \hskip-2emf(t)=\sum _{k=-1}^{\infty }b_{k} \exp \left(-\lambda _{k} t\right) =
-\frac{1}{J(-B_{\alpha } )}
\exp \left(-B_{\alpha } t\right)+\\
&&\hskip3em + \frac{1}{B_{\alpha }
-p^{-\alpha } - (1-p^{-\alpha } )\delta _{0} } u_{k} \exp
\left[-(p^{-\alpha } + (1-p^{-\alpha } )\delta _{0}
)t\right]+g(t) ,
\end{eqnarray*}
where we have set
$$
 g(t)\equiv
\sum _{k=1}^{\infty }\frac{1}{B_{\alpha } -p^{-\alpha (k+1)}
-p^{-\alpha k} (1-p^{-\alpha } )\delta _{k} } u_{k} \exp
\left[-(p^{-\alpha (k+1)} +p^{-\alpha k} (1-p^{-\alpha } )\delta
_{k} )t\right]  .
$$
For   $g(t)$, we have the estimate
$$
 \frac{1}{B_{\alpha } }
\sum _{k=1}^{\infty }u_{k} \exp \left[-p^{-\alpha k} t\right]
<g(t)<\frac{1}{B_{\alpha } -p^{-\alpha } } \sum _{k=1}^{\infty
}u_{k} \exp \left[-p^{-\alpha k} p^{-\alpha } t\right]  .
$$
Further, taking into account \eqref{(B.6)}--\eqref{(B.8)},
we find that
  \begin{eqnarray}
&& \hskip-5em \frac{\tilde{D}(\alpha )}{B_{\alpha } } \sum
_{k=1}^{\infty }p^{(1-2\alpha )k} \left(1+o(1)\right)\exp
\left[-p^{-\alpha k} t\right]< g(t)<\nonumber\\
&&< \frac{\tilde{U}(\alpha )}{B_{\alpha } -p^{-\alpha } }
\sum _{k=1}^{\infty }p^{(1-2\alpha )k} \left(1+o(1)\right)\exp
\left[-p^{-\alpha k} p^{-\alpha } t\right],\quad \alpha>1; \label{(B.9)}\\[10pt]
&&  \hskip-5em \frac{\tilde{D}(\alpha )}{B_{\alpha } }
\sum _{k=1}^{\infty }p^{-k} \left(1+o(1)\right)\exp
\left[-p^{-\alpha k} t\right] <g(t)<\nonumber \\
&&< \frac{\tilde{U}(\alpha
)}{B_{\alpha } -p^{-\alpha } } \sum _{k=1}^{\infty }p^{-k}
 \left(1+o(1)\right)\exp \left[-p^{-\alpha k} p^{-\alpha } t\right]   ,
 \quad \alpha <1;     \label{(B.10)}\\[10pt]
&&  \hskip-5em \frac{\tilde{D}(1)}{B_{1} }
\sum _{k=1}^{\infty }\frac{p^{-k} }{k^{2} } \left(1+o(1)\right)
\exp \left[-p^{-k} t\right] <g(t)<\nonumber\\
&& <\frac{\tilde{U}(1)}{B_{1}
-p^{-1} }
 \sum _{k=1}^{\infty }\frac{p^{-k} }{k^{2} } \left(1+o(1)\right)\exp
  \left[-p^{-k} p^{-1} t\right],\quad\alpha=1. \label{(B.11)}
   \end{eqnarray}
From \eqref{(B.9)}--\eqref{(B.11)}, using the inequalities \eqref{(A.4)}, \eqref{(A.2)},
 we obtain  \eqref{(2.19)}--\eqref{(2.21)}.
The proof is complete.


\subsection*{Acknowledgments}
The authors wish to express their gratitude to Prof. Igor Volovich,
Nikolai Shamarov, and Alexei Dolgov  for useful discussions.
\bigskip

This work has been partially supported by the RFBR (grants No:
05-03-32563a, 07-02-00612a) and the Program OCHNM RAS (1-OCH/06-08).


\end{document}